\documentclass[12pt]{scrartcl}
\usepackage[american]{babel}
\usepackage{graphicx}
\usepackage{natbib}
\usepackage{graphicx}
\usepackage{amssymb,amsmath}
\usepackage{booktabs}
\usepackage{longtable}
\usepackage{multirow}
\usepackage{authblk}
\usepackage{xcolor}
\LTcapwidth=\textwidth

\def\spacingset#1{\renewcommand{\baselinestretch}%
{#1}\small\normalsize}

\date{}
\author{Louis Raynal\thanks{Equal contribution.}\protect\phantom{\footnotesize *}\thanks{ORCID: 0000-0003-2805-3254}}
\author{Till Hoffmann\protect\footnotemark[1]\protect\phantom{\footnotesize *}\thanks{E-mail: thoffmann@hsph.harvard.edu $\cdot$ ORCID: 0000-0003-4403-0722}}
\author{Jukka-Pekka Onnela\thanks{E-mail: onnela@hsph.harvard.edu $\cdot$ ORCID: 0000-0001-6613-8668}}
\affil{\small Department of Biostatistics, T.H. Chan School of Public Health, Harvard University}

\newcommand{\supref}[2]{#2}

\usepackage[colorlinks, allcolors=blue]{hyperref}
\usepackage{hyperref}
\usepackage{dsfont} 

\newcommand{\parenth}[1]{\left(#1\right)}
\newcommand{\braces}[1]{\left\{#1\right\}}

\begin{document}

\title{Cost-based feature selection for network model choice}

\maketitle

\begin{abstract}

Selecting a small set of informative features from a large number of possibly noisy candidates is a challenging problem with many applications in machine learning and approximate Bayesian computation. In practice, the cost of computing informative features also needs to be considered. This is particularly important for networks because the computational costs of individual features can span several orders of magnitude. We addressed this issue for the network model selection problem using two approaches. First, we adapted nine feature selection methods to account for the cost of features. We show for two classes of network models that the cost can be reduced by two orders of magnitude without considerably affecting classification accuracy (proportion of correctly identified models). Second, we selected features using pilot simulations with smaller networks. This approach reduced the computational cost by a factor of 50 without affecting classification accuracy. To demonstrate the utility of our approach, we applied it to three different yeast protein interaction networks and identified the best-fitting duplication divergence model. 
Computer code to reproduce our results is available online.

\end{abstract}

\noindent%
\textit{Keywords:} feature selection, mechanistic network models, classification, cost-based feature selection, approximate Bayesian computation

\spacingset{1.5} 

\section{Introduction}

As data grow richer and more complex, it is becoming ever more important to select informative features from a large number of candidates to address various inferential problems. This challenge is pervasive in science and affects numerous fields, including population genetics \citep{pritchard:etal:1999}, particle physics \citep{brehmer2018}, and cosmology \citep{Alsing2019}.
Dimensionality reduction to analyze complex datasets spans diverse domains and is a necessary step in a wide variety of statistical methods, such as regression models \citep{langley:sage:1997}, machine learning \citep{blum:langley:1997}, and approximate Bayesian computation (ABC) \citep{blum:2010:sumstat,sisson:etal:2019}. In practice, however, it is important to select not only informative features but also features that have low computational cost to facilitate fast model iteration and analysis.

Selecting features for networks is particularly challenging. For example, we cannot usually rely on (conditional) independence assumptions that may be suitable for survey data or biological assays because different parts of the network may be interdependent. Network models can be assigned to two broad classes: statistical network models and mechanistic network models \citep{Goyal2020}. While inference for statistical network models is generally feasible because the likelihood can be evaluated or approximated, learning from data using mechanistic network models is difficult because the likelihood is often intractable \citep{chen:etal:2019}. A mechanistic network model is defined in terms of a rule set that is repeatedly applied to a network starting from a seed graph~\citep{Overgoor2019}, where adding or removing nodes or edges is conditional on the current state of the graph. For example, the Barabási–Albert (BA) model \citep{barabasi:albert:1999} connects a new node to existing nodes at each step with probability proportional to their degree. For protein interaction networks, duplication divergence models grow a network by duplicating existing nodes at each step of the growth process and randomly adding or deleting edges~\citep{sole:etal:2002,vazquez:etal:2003}. Because the order of addition of nodes is typically unknown, the likelihood is intractable. We thus need to resort to likelihood-free inference techniques, such as ABC \citep{sisson:etal:2019}.

Because we are interested in mechanistic network models, we seek strategies to address the computational bottleneck that can occur when computing network features. In this setting, we explored two strategies to select features efficiently. First, we focused on feature selection methods that take into account each feature's computational time. These \textit{cost-based} methods identify a set of features that are fast to compute yet informative. Second, to reduce the computational cost of the selection process, we evaluated features for smaller graphs with fewer nodes than the observed graph. While the first strategy is more general, the second strategy applies only to mechanistic network models. We investigated these approaches using two simulated classification problems, one to discriminate between two different BA models \citep{barabasi:albert:1999} and the other to discriminate between two protein interaction network models (the duplication mutation complementation model \citep{vazquez:etal:2003} and the duplication with random mutation model \citep{sole:etal:2002}). We also used the features selected by our methods to identify the best-fitting network model for three yeast protein interaction datasets.

In this paper, we first describe the model choice setting in Section \ref{sec:matAndMet}, including the generation of a reference table and the different methods that will be used. We then detail the development of a suite of cost-based feature selection methods in Section~\ref{sec:cost-based-methods} and consider the feasibility of selecting informative features using pilot simulations of networks with fewer nodes in Section~\ref{sec:pilot-simulations}. In Section~\ref{sec:experiments}, we describe the validation of both approaches using simulation and the application of our methods to identify the best-fitting model for three different yeast protein interaction networks. Finally, we  discuss the results and consider opportunities for future research in Section~\ref{sec:discussion}.

\section{Materials and methods}
\label{sec:matAndMet}

We considered a problem with $m$ (mechanistic network) models and investigated which model would most likely give rise to the observed data $z$. The parameter of interest was the model index $y$, which we assigned a categorical prior. Each model had model-specific parameters $\theta^{(y)}$ and a corresponding prior. Given $q$ candidate features $x\parenth{z}=\braces{x_1\parenth{z},\ldots,x_q\parenth{z}}$, our objective was to select a subset of $q'\ll q$ features that can discriminate between different models. To this end, we simulated a reference table of size $N$ comprising candidate features in four steps. First, we sampled the model index $y$ from the categorical prior. Second, we sampled parameters $\theta^{(y)}$ from the model-specific prior. Third, we generated a synthetic dataset $z$ using model $y$ and parameters $\theta^{(y)}$. Finally, we evaluated the candidate features $x(z)$ on the synthetic dataset $z$. In other words, we evaluated candidate features on samples drawn from the prior predictive distribution. When considering the model indices as responses and features as predictors, the table can be used to train a supervised classifier that  predicts the model index for any set of features corresponding to a new observation. 

Feature selection methods belong to three broad classes: \textit{filter} methods, \textit{embedded} methods, and \textit{wrapper} methods \citep{jovic:etal:2015}.
A \textit{filter} method is a preprocessing feature selection step, independent of the learning algorithm. It determines the relevance of features using various measures such as correlation or mutual information \citep{shannon:1948} rather than classification accuracy.
A \textit{wrapper} directly uses the learning algorithm to evaluate the relevance of feature subsets in terms of classification accuracy.
An \textit{embedded} approach performs feature selection during the training of the learning algorithm, as in the case of random forests (RF) \citep{breiman:2001} or LASSO \citep{tibshirani:1996}. We focused on filter methods because they can be used as a preprocessing step before applying other selection techniques, and they are faster than wrapper and embedded methods \citep{Bolon-Canedo2013}. Unless otherwise specified, we use feature selection methods to refer to filter methods for feature selection.

Studying large network datasets can be computationally challenging depending on the complexity of the features. To address this problem, we investigated two independent strategies: feature selection methods that account for the relative computational cost of candidate features  (Section~\ref{sec:cost-based-methods}) and pilot simulations of smaller networks to select features (Section~\ref{sec:pilot-simulations}).

\subsection{Cost-based feature selection methods}
\label{sec:cost-based-methods}

When two features are equally informative for the model selection task, the one that is computationally less expensive is preferred. Ultimately, we seek to balance informativeness and computational cost; here, we describe the development of a suite of \emph{cost-based} feature selection methods to achieve this goal.

The first applications of such cost-based methods for feature selection were in medicine \citep{Bolon-Canedo2015}, where a feature carries information about a patient but retrieving the feature carries a financial cost depending on its nature (e.g., PET scanning is significantly more expensive than X-ray imaging). The objective is to achieve a trade-off between the financial cost to obtain the feature and the informativeness of the feature for the classification problem.
\citet{bolon-canedo:etal:2014} proposed an adaptation of the ReliefF algorithm \citep{kononenko:1994}. A more general framework that adapts the correlation-based and minimal-redundancy-maximal-relevance criteria was developed by \citet{peng:etal:2005, bolon-canedo:etal:2014:framework}. \citet{zhou:etal:2016} made use of \citet{breiman:2001}'s RF algorithm where the feature cost is used at each internal node of a classification tree to sample inexpensive features more often than expensive ones.

To apply cost-based methods in our feature selection framework, we first replaced the financial cost of a feature by the computational cost required to evaluate the feature. We estimated the time required to evaluate each feature during the generation of the reference table, and we averaged the times over all simulations to obtain an average cost vector $C=\braces{C_1, \ldots, C_q}$. We then rescaled the cost vector by dividing it by its sum, such that the rescaled cost is the proportion of computational cost incurred by evaluating the corresponding feature.

Some of the features could be computed in groups to economize resources. For example, computing multiple characteristics of the degree distribution through a single evaluation of the degree density reduces the per-feature computational cost. However, the cost remains unchanged even if just one characteristic is computed. In other words, dividing the cost by the number of characteristics in the group deflates the cost of each characteristic; if only one of the characteristics is selected, one would nevertheless incur its full cost. To avoid bias, we assigned the full computational cost to each feature (i.e., the cost incurred if computed independently) even for features that could be computed in groups.

We used the nine feature selection methods and their possible cost-based versions presented below. These methods belong to three broad categories: methods based on mutual information \citep{shannon:1948}, the ReliefF algorithm \citep{kira:rendell:1992, urbanowicz:etal:2018:review}, and RF feature importance \citep{breiman:2001}. We describe the benefits of each method below. 

\subsubsection{Mutual information-based approaches}

\citet{shannon:1948}'s mutual information (MI) is an information theoretic measure that quantifies how much knowledge of one random variable reduces uncertainty in another, i.e., it quantifies the amount of information one variable carries about another. MI can capture non-linear dependencies between variables and is invariant under invertible and differentiable transformations of the variables \citep{kraskov:etal:2004, cover:thomas:2012}. These advantages make it popular among feature selection methods \citep{brown:etal:2012, vergara:estevez:2014}.

Given two arbitrary discrete random variables $X_1$ and $X_2$ with values in the sets $\mathcal{X}_1$ and $\mathcal{X}_2$, respectively, their mutual information is
\begin{equation*}
    I(X_1;X_2) = \sum_{x_1 \in \mathcal{X}_1} \sum_{x_2 \in \mathcal{X}_2} p(x_1, x_2) \log\left( \frac{p(x_1, x_2)}{p(x_1)p(x_2)} \right),
\end{equation*}
where $p(\cdot)$ and $p(\cdot,\cdot)$ denote the marginal and joint probability mass functions, respectively. 
Similarly, the conditional mutual information $I(X_1; X_2 \mid X_3)$ measures the information between two features given knowledge of a third $X_3\in\mathcal{X}_3$. It is defined as 
\begin{equation*}
    I(X_1; X_2 \mid X_3) = \sum_{x_3 \in \mathcal{X}_3} p(x_3) I(X_1; X_2 \mid X_3=x_3),
\end{equation*}
where $I(X_1; X_2 \mid X_3=x_3)$ is the mutual information evaluated on the subset satisfying $X_3=x_3$. For the purposes of the experiments in Section~\ref{sec:experiments}, we used an adjusted mutual information estimator to account for chance agreement~\citep{vinh:2010}.

All of the following feature selection methods based on MI are sequential forward methods, i.e., the feature subset is empty at the start and features are sequentially selected and added to it one at a time. In other words, at the $k^\text{th}$ step, $k-1$ features are already included, and the newly selected $X_k$ is the feature that maximizes a quality function $J(X_k)$. Here, we describe three methods based on different information-theoretic quality measures that are commonly used. They are of particular interest because they can avoid redundant features and identify positive interactions between features \citep{brown:etal:2012}.

\paragraph{Minimal-redundancy-maximal-relevance}

The minimal-redundancy-maximal-rel\-e\-vance (mRMR) method \citep{peng:etal:2005} relies on the quality function
\begin{equation}
    J_\text{mRMR}(X_k) = I(X_k ; y) - \sum_{j \in \mathcal{S}} I(X_j ; X_k),
\label{eq:mRMR:crit}
\end{equation}
where $\mathcal{S}$ designates the set of feature indices already selected and $y$ is the model label.
As its name suggests, the first term quantifies the relevance of feature $X_k$ for identifying the model $y$, and the second corresponds to the redundancy between a new candidate feature $X_k$ and the features already selected $\mathcal S$.

\paragraph{Joint mutual information}
\label{subsubsec:joint-mutual-information}

\citet{yang:moody:1999} proposed to use the joint mutual information (JMI)
\[
I( (X_j,X_k) ; y)=\sum_{x_1\in\mathcal{X}_1}\sum_{x_2\in\mathcal{X}_2}\sum_{y\in\mathcal{y}}p\parenth{x_1,x_2,y}\log\parenth{\frac{p\parenth{x_1,x_2,y}}{p\parenth{x_1,x_2}p\parenth{y}}}.
\] 
JMI quantifies the information between a pair of features $\parenth{X_j,X_k}$ and the model label $y$.
The quality function applied to a candidate feature $X_k$ is
\begin{equation}
    J_\text{JMI}(X_k) = \sum_{j \in \mathcal{S}} I( (X_j,X_k) ; y).
\label{eq:JMI:basic}
\end{equation}
Interestingly, maximizing Equation \eqref{eq:JMI:basic} is equivalent to maximizing \citep{brown:etal:2012}
\begin{equation}
    I(X_k ; y) - \sum_{j \in \mathcal{S}} I(X_j ; X_k)
    + \sum_{j \in \mathcal{S}} I(X_j ; X_k \mid y)
\label{eq:JMI:alternative}
\end{equation}
which is the mRMR criterion in Equation \eqref{eq:mRMR:crit} plus a positive term for the MI between the already selected features $\mathcal S$ and $X_k$ conditional on $y$. Due to the inclusion of this term, JMI is able to detect positive interactions between features with respect to the model $y$.
As highlighted by \citet{brown:etal:2012}, 
while the terms $I(X_j;X_k)$ are negative to reduce the redundancy between features, the conditional MI terms are positive. Thus ``the inclusion of correlated features can be useful, provided the correlation within classes is stronger than the overall correlation'' \citep{brown:etal:2012}.

\paragraph{Joint mutual information maximization}

\citet{bennasar:etal:2015} proposed a ``maximization of the minimum'' (maximin) approach by replacing the sum in Equation \eqref{eq:JMI:basic} with the minimum over all previously selected features. The criterion is
\begin{equation*}
    J_\text{JMIM}(X_k) = \min_{j \in \mathcal{S}}  I( (X_j,X_k) ; y).
\end{equation*}
Similarly to $J_\text{JMI}(\cdot)$, maximizing this criterion is equivalent to maximizing
\begin{equation}
    \min_{j \in \mathcal{S}} \left[ I(X_k ; y) + I(X_j ; y) - I(X_k ; X_j)
    + I(X_k ; X_j \mid y) \right].
\label{eq:JMIM:alernative}
\end{equation}
While the alternative expressions in Equations \eqref{eq:JMI:alternative} and \eqref{eq:JMIM:alernative} may seem more complicated, they are practical as they do not require the evaluation of MI between the model label and a pair of features; rather, they only require (conditional) MI between univariate quantities.

\paragraph{Cost-based versions}

It is easy to construct cost-based versions of these three feature selection methods by penalizing the quality measure $J\left(X_k\right)$ of each feature by its cost $C_k$. In other words, the modified quality function is
\begin{equation*}
    J(X_k) - \lambda C_k,
\end{equation*}
where $\lambda$ is a positive parameter that balances the relevance of a feature and its cost. For this investigation, we will use the mRMR adaptation that was introduced by \cite{bolon-canedo:etal:2014}, as well as our cost-based adaptations of JMI and JMIM. These two methods are especially interesting for their potential to handle positive interactions between features.

\subsubsection{Random forest importance-based approaches}

Next, we considered feature selection methods methods based on \citet{breiman:2001}'s random forest (RF) measures of importance to obtain a ranking of features. RFs are supervised learning algorithms with a feature ranking technique. In the ABC setting, these methods have been used for model choice \citep{pudlo:etal:2016} and parameter inference \citep{raynal:etal:2019}; here, we applied them to feature selection.

A RF is an ensemble of decision trees \citep{breiman:etal:1984}, the construction of which is randomized by using bootstrap samples for each tree and subsampling the covariates at each tree node. A decision tree is built by sequentially partitioning the covariate space according to a covariate and a split value so that the cut maximizes an information gain criterion, such as the gini impurity \citep{biau:scornet:2016}. The criterion is maximized only over a subset of the features; $\sqrt{q}$ is a common default choice in classification and was used in our analysis. 

RFs can be used to rank covariates based on their relevance for the learning task.
Two measures of feature importance are commonly used: the mean decreased impurity (MDI) and the mean decreased accuracy (MDA) \citep{biau:scornet:2016}. 
To compute the MDI for a given feature, the gini impurity can be summed over all trees and all nodes where this feature has been used.
The MDA of a feature is computed over all trees as the decrease in accuracy obtained on out-of-bag data when randomly permuting its covariate values. Out-of-bag data refers to data points that were not selected in a given bootstrap sample and thus were not used to construct the given tree. For this investigation, we examined two alternative cost-based variants. The first is a generalization of the proposal by \citet{zhou:etal:2016}, and the second is a simple penalization of the MDI or MDA importance measures.

\paragraph{Weighted random forest adaptation}

In the standard RF feature selection algorithm, the set of features considered at each internal node are drawn uniformly at random.
\citet{zhou:etal:2016} proposed training a RF by replacing the uniform sampling of covariates at each node of each tree by weighted sampling such that expensive features are less likely to be selected. The RF then uses the resulting feature ranking (measured with the MDI or MDA) to determine which features to retain. The sampling weights are defined by the reciprocal of their cost, so that a feature $k$ has sampling probability $w_k\propto C_k^{-1}$.

We considered a generalization of this idea to generate sampling weights that depend on a tuning parameter $\lambda$ similar to the other cost-based methods examined in this paper. In particular, we used $w_k\propto C_k^{-\lambda}$ such that the importance of feature cost increases with $\lambda$. This approach allows a smooth transition between a classic (i.e., not cost-based) RF algorithm ($\lambda=0$) and the strategy of \citet{zhou:etal:2016} ($\lambda=1$).
While $\lambda$ influences the cost of other methods through a multiplicative term, the exponentiation here has a much greater impact on the final weights as $\lambda$ grows.

\paragraph{Penalized random forest feature importance}

We also used a more naive cost-based approach by penalizing the RF importance values from a forest built on a simulated reference table. In this method, we retrieve the feature importance measures (MDI or MDA) and, after normalization between zero and one, subtract $\lambda C_k$ from each corresponding $k^\text{th}$ feature importance. Normalization makes the two measures more comparable to each other for the same value of $\lambda$. This approach is similar to the cost-based ReliefF algorithm because the relevance scores are simply penalized by the feature cost.

\subsubsection{ReliefF-based approaches}
\label{par:relief-approaches}

The final class of feature selection methods we consider are ReliefF methods \citep{kononenko:1994}, a multi-class extension of the original Relief algorithm \citep{kira:rendell:1992}. Relief determines a weight (or score) for each feature that increases with its importance. Relief-based methods quantify the relevance of a feature based on how well it separates data with different labels in the feature space and how close data with identical labels are to each other. Because the feature weights are updated based on nearest neighbors, the weights indirectly depend on the whole feature space; thus, this method is capable of detecting interactions among features \citep{urbanowicz:etal:2018:review}. For this investigation, ReliefF was also of interest because of existing research on cost-based versions of the method \citep{bolon-canedo:etal:2014}.

At each iteration of the ReliefF algorithm, a labeled data point $R^{(i)}:=(y^{(i)}, x^{(i)})$ (a.k.a., instance) is randomly selected without replacement and its $\ell$ nearest neighbors within each class are identified. The neighbors of $i$ within the same class $H^{(i)}$ are cal\-led ``hits'' and those in different classes $M^{(i)}$ are cal\-led ``mis\-ses.''
Each feature weight is updated based on the following rule:
A feature that is relevant for distinguishing between classes should increase the distance between $R^{(i)}$ and its misses, while the distances between $R^{(i)}$ and its hits should be small.
The former requirement yields a positive contribution to the importance weight because the feature can separate different classes; the latter penalizes the feature score if members of the same class are too far apart.

The algorithm cycles $r$ times through the process of selecting a random instance $R^{(i)}$ and updating weights. 
In our experiments, as proposed by \citet{urbanowicz:etal:2018:benchmark}, each training instance was selected successively (i.e., $r=N$ without randomization). Following \citet{urbanowicz:etal:2018:review}, we used $\ell=10$ for the number of nearest neighbors within each class. The distance between two instances $R^{(i)}$ and $R^{(j)}$ on the $k^\text{th}$ covariate dimension is
\begin{equation}
    {
    d_{k}(R^{(i)}, R^{(j)}) = \begin{cases}
    \frac{\mid x_k^{(i)} - x_k^{(j)} \mid }{\max(X_k) - \min(X_k)}  & \text{if $X_k$ is continuous}\\
    \mathds{1}_{ \{ x_k^{(i)} \neq x_k^{(j)} \} } & \text{if $X_k$ is discrete}
  \end{cases}
  }
  \label{eq:relief-distance}
\end{equation}
where $\mathds{1}$ denotes the indicator function. This distance is summed over all dimensions to determine the nearest neighbors.

\citet{bolon-canedo:etal:2014} proposed a cost-based adaptation of ReliefF by penalizing the weight update expression, leading to
\begin{multline}
    w_k \rightarrow w_k - \frac{1}{r \times \ell}\left\{\sum_{j\in H^{(i)}} d_k\left(R^{(i)},R^{(j)}\right)\right. \\ 
            \left. + \sum_{j\in M^{(i)}} \left[ \frac{p(y^{(j)})}{1-p(y^{(i)})}
            \times d_k(R^{(i)}, R^{(j)}) \right]\right\} - \frac{\lambda C_k}{r}.
    \label{eq:reliefF-weight-update}
\end{multline}
We used $\frac{\lambda C_k}{r}$ instead of $\frac{\lambda C_k}{r\ell}$ as originally proposed \citep{bolon-canedo:etal:2014} to ensure that the impact of $\lambda$ on feature rankings does not depend on the number of nearest neighbors $\ell$. The cost penalty is applied independently for each instance, and penalizing the weights of a classic ReliefF algorithm by $\lambda C_k$ is equivalent.

One issue with ReliefF is its sensitivity to noise features, which can change the identity of the nearest neighbors and thus the final feature scores and ranking, especially given the range normalization in Equation \eqref{eq:relief-distance}.
To prevent this issue and possibly improve the quality of the selected features, we used \citet{breiman:2001}'s proximity matrix to identify similar instances. The proximity matrix is obtained by training a RF using the model indices as responses and the features as predictors. Its entries denote the average number of times two data points fall in the same leaf, where the average is taken over all trees. This similarity measure has the advantage of being based mostly on relevant features. To guarantee enough data with non-zero similarities, we built shallow trees with a minimum leaf size of 100 instances.

\subsection{Feature selection using small pilot simulations}
\label{sec:pilot-simulations}

The inferential scheme of this paper includes two reference tables, one for selecting an informative subset of features and one for evaluating the selected features. Both are obtained by simulating networks with $n$ nodes, the number of nodes in the observed network. Recall that mechanistic models generate networks by sequentially adding nodes to a small seed graph according to a given mechanism until $n$ nodes are present. For large networks, generating the first table can be very expensive. We reduced the generation time by reducing the number of nodes of the simulated networks. In particular, we simulated smaller pilot networks with $n_s<n$ nodes to obtain the reference table for feature selection. We then investigated whether the features selected to classify smaller networks remained relevant for classifying the larger networks. Our strategy of using smaller networks can be interpreted as stopping network construction early to assess the ability of the associated features to distinguish between the different models.

We implemented both the cost-based feature selection methods presented above and feature selection based on pilot simulations in a Python package (see Section~\ref{sec:supplement}).

\section{Experiments}
\label{sec:experiments}

\subsection{Cost-based feature selection}
\label{sec:efficiency_cost_based_selection}

To study the efficacy of the feature selection methods presented in Section \ref{sec:matAndMet}, we considered two problems of network model choice: classifying four different Ba\-ra\-bá\-si–Al\-bert (BA) models \citep{barabasi:albert:1999} and distinguishing between two protein interaction network models \citep{sole:etal:2002,vazquez:etal:2003}.

\subsubsection{Barabási–Albert models}
\label{subsec:BA:cost_based_1000nodes}

The BA model is a simple yet influential mechanistic network model of undirected networks with two parameters. Starting with a small seed graph, a new node is added to the network at each time step until the number of nodes reaches $n$. Each new node is connected to $m$ existing nodes with linear preferential attachment, i.e., neighbors are selected with probability proportional to their degree. As a simple test case, we considered a four-class classification problem based on the BA model by considering four different values of the number of neighbors $m\in\left\{1,2,3,4\right\}$ while keeping the number of nodes fixed at $n=1,000$. We used the package \texttt{networkx} \citep{hagberg:etal:2008} to generate BA networks starting with a star graph with $m + 1$ nodes.

We generated $N=5,000$ network realizations using a uniform prior for the number of neighbors $m$ (equivalently, model labels). A reference table of candidate features was obtained by evaluating 51 statistics on the simulated networks (see \supref{Table \ref{sup:tab:feature-statistics-used} in Appendix \ref{sup:sec:feature_statistics}}{Table~1 of the online supplement} for details). These features comprise a wide variety of network features as well as four inexpensive noise covariates that represent independent realizations from different distributions: standard normal, Bernoulli with success probability 0.5, and continuous and discrete uniform, both on the interval $[0,50]$. Such noise features can be used to investigate whether a feature selection method can correctly discard spurious features.

Because of the simplicity of the models, it is easy to discriminate among them. On the one hand, certain features (such as the mean degree) can uniquely identify each model. On the other hand, some features (such as the number of nodes in the largest connected component) did not vary across simulations. We therefore discarded the latter before applying feature selection methods.

Using each of the feature selection algorithms, we ranked features for a range of cost regularization values $\lambda$. We evaluated the feature rankings in three steps. First, we generated a second, independent reference table of the same size $N=5,000$ as an independent test set. Second, we selected the top $k$ features for each ranking, varying $k$ between 1 and 15 features. Finally, we trained a nearest-neighbor classifier with 10 neighbors (after standardizing features) and evaluated the classification accuracy (i.e., the fraction of correctly identified models) using 10-fold cross-validation on the test set. We used a nearest-neighbor classifier because its prediction can be understood as the maximum a posteriori estimate under ABC \citep{biau:etal:2015}. We also considered a support vector machine classifier, but it had lower accuracy throughout. 

\begin{figure}
    \spacingset{1}
    \centering
    \includegraphics{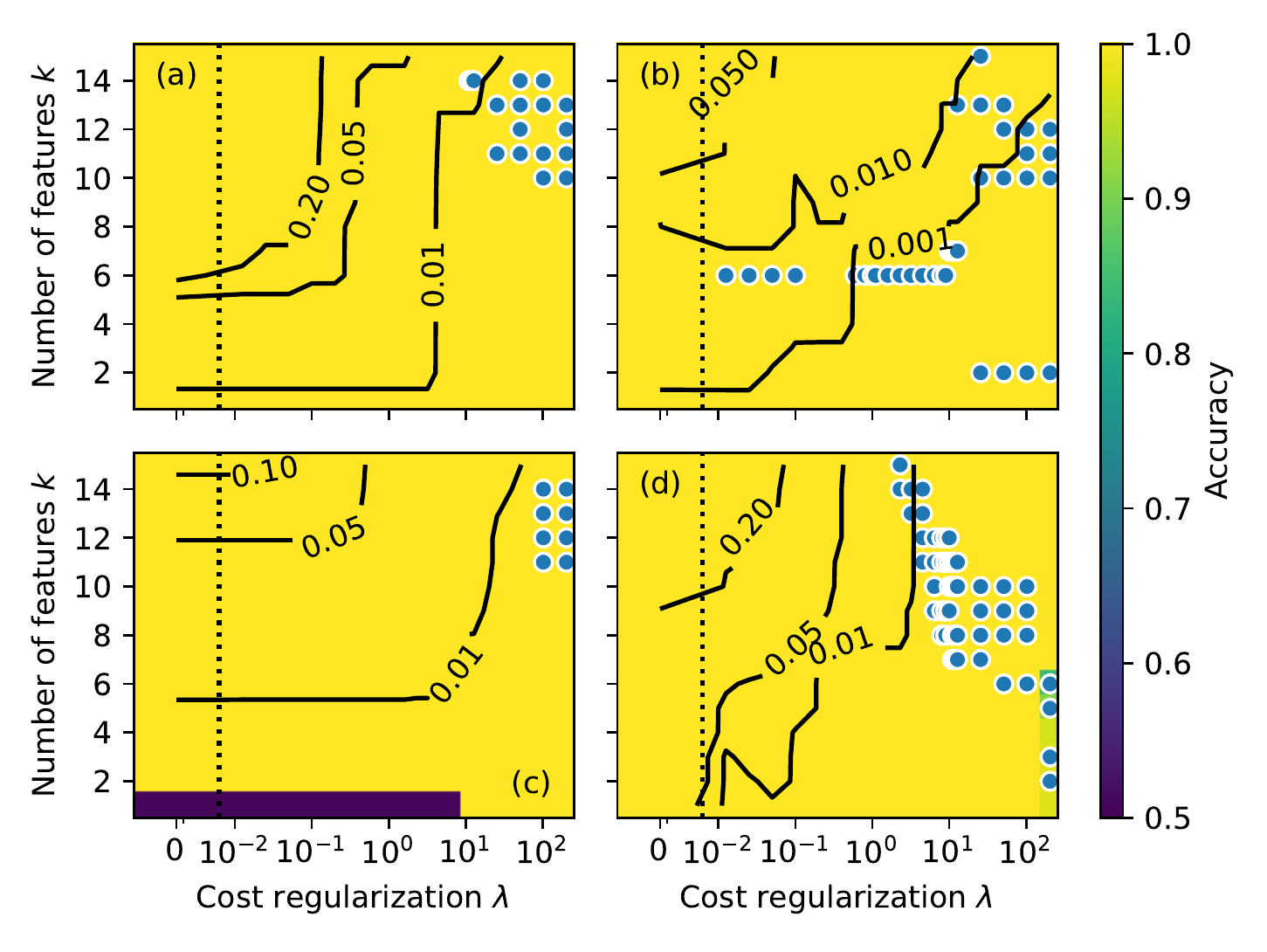}
    \caption{\emph{Cost-based feature selection can reduce computational cost by more than two orders of magnitude without affecting accuracy.} Each panel shows the classification accuracy of different BA models as a heatmap for different cost regularizations $\lambda$ and number of selected features $k$. Contours indicate the relative cumulative cost for the selected features, i.e., the fraction of computational cost that could be saved by using the feature subset compared with using all features. Markers indicate noise features that tended to be selected for large cost regularization. Panels~(a) through~(d) show results for the JMI, mRMR, distance-based reliefF, and penalized random forest importance methods based on mean decreased accuracy, respectively. While we used the same grid of regularization values $\lambda$, they may have a different effect for different methods. The dashed vertical lines separate $\lambda=0$ to the left from $\lambda>0$ to the right.}
    \label{fig:ba-accuracy-matrix}
\end{figure}

As intended, the total cost of the selected features decreased substantially with increasing cost regularization $\lambda$. As shown in Figure \ref{fig:ba-accuracy-matrix} for four methods, cost-based feature selection can reduce the computational cost by more than two orders of magnitude without affecting accuracy compared with using all features. Corresponding figures for the other methods can be found in \supref{Figure~\ref{fig:ba-accuracy-matrix-appendix} in Appendix~\ref{sup:sec:heatmaps}}{Figure~1 of the online supplement}. Large values of $\lambda$ led to the inclusion of inexpensive yet uninformative noise features. 

Information-theoretic approaches (JMI, JMIM, and mRMR) perfectly classified different BA models for all combinations of cost regularizations and number of selected features. As shown in panel~(b) of Figure~\ref{fig:ba-accuracy-matrix}, mRMR selected noise features earlier than other methods because they have minimal redundancy with other features (cf. Equation~\eqref{eq:mRMR:crit}). RF-based approaches exhibited some variability in feature rankings because different bootstrap samples used in the construction of the forest can result in different feature scores. However, irrespective of the specific ranking, RF-based methods performed well except for very large cost regularization values. As shown in panel~(d) of Figure \ref{fig:ba-accuracy-matrix}, accuracy decreased because the strong cost regularization led to the inclusion of a large number of noise features. Cost-based ReliefF methods performed well if more than two features were selected. However, they prioritized a weakly informative feature, the size of the three-core, such that classification based on a single feature performed poorly. 

This is a fundamental limitation of ReliefF that is not related to our cost-based adaptation. In particular, the three-core of a network is the maximal subgraph induced by nodes that have at least three neighbors \citep{Batagelj:2003}, i.e., the three-core is defined recursively. The three-core is empty for BA models with $m<3$, and it includes all nodes for BA models with $m\geq 3$. The size of the three-core is thus a perfect feature for discriminating between $m<3$ and $m\geq 3$, but it cannot discriminate between $m=1$ and $m=2$ or between $m=3$ and $m=4$. Nevertheless, ReliefF assigns a higher score to the three-core feature than, for example, the mean degree, which can easily distinguish all four models. Specifically, the within-class distance in Equation \ref{eq:reliefF-weight-update} for both features is zero. However, after range normalization, the mean out-of-class distance for the three-core is $2/3$ whereas the mean out-of-class distance for the mean degree feature is only $5/9$; thus, the size of the three-core is ranked higher despite being less informative (see \supref{Appendix~\ref{app:three-core}}{Section~3 of the online supplement} for details). The accuracy increases with larger cost regularization as the relatively expensive three-core feature is replaced by cheaper and more informative features.

The results of this experiment demonstrate that it is possible to decrease the total feature cost by more than two orders of magnitude without decreasing the classification accuracy. We consider a more difficult scenario below.

\subsubsection{Models for protein interaction networks}
\label{subsec:DMC_DMR:cost_based_1000nodes}

To further evaluate the various cost-based feature selection methods, we next focused on the classification of two mechanistic network models commonly used to describe protein interaction networks: the duplication mutation complementation (DMC) \citep{vazquez:etal:2003} and duplication with random mutation (DMR) \citep{sole:etal:2002} models.

For both models, each step of network generation starts by adding a new node to the network and connecting it to all neighbors of a seed node chosen uniformly at random, i.e., a random seed is duplicated. For the DMC model, for each neighbor of the duplicated node, either the edge to the duplicated node or the edge to the new node (but never both) is removed with probability $\theta_\text{del}^\text{(DMC)}$. Finally, an edge between the new and duplicated node is added with probability $\theta_\text{conn}^\text{(DMC)}$.
For the DMR model, only the edges adjacent to the new node are erased with probability $\theta_\text{del}^\text{(DMR)}$, and edges between the new and any other node in the network are added with probability $\theta_\text{new}^\text{(DMR)} / n(t)$, where $n(t)$ denotes the number of nodes in the network at the beginning of step $t$. These rules are applied repeatedly until the desired number of nodes $n=1,000$ is reached.

Using a pair of connected nodes as the seed graph, we generated a reference table comprising $N=5,000$ simulations by choosing each of the two models with equal probability and evaluating the features as described above. The priors on parameters were independent uniform distributions on the interval $[0.25, 0.75]$ for each parameter. We chose these bounds to avoid unlikely graphs that were either under- or over-connected. A second reference table of identical size was generated and used as a validation set using 10-fold cross-validation as described above.

\begin{figure}
    \spacingset{1}
    \centering
    \includegraphics{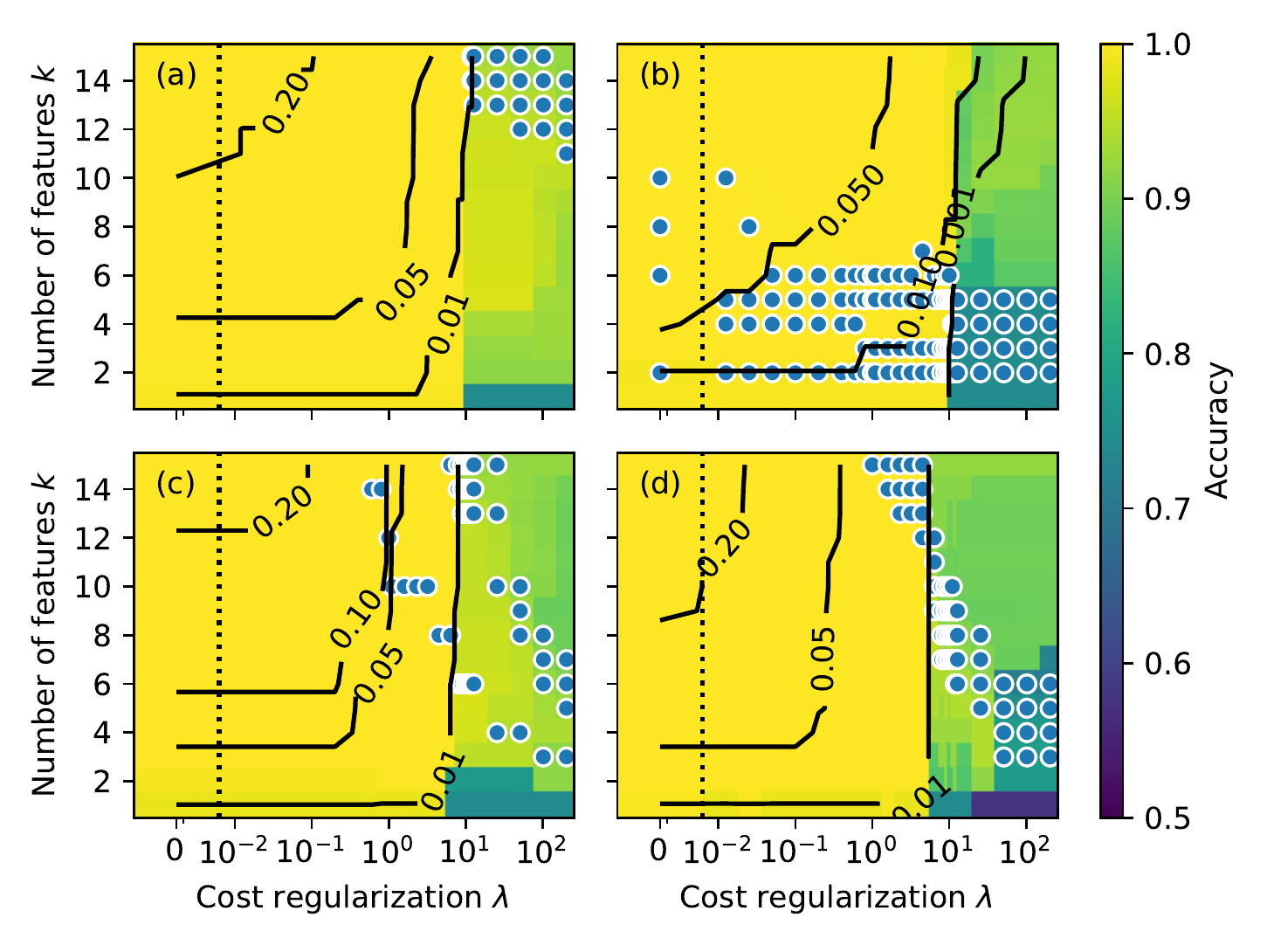}
    \caption{\emph{The identification of duplication divergence models for protein interaction networks based on cost-based feature selections exhibits a sharp transition as the cost regularization increases.} Each panel shows the classification accuracy of different duplication divergence models as a heatmap. See the caption of  Figure~\ref{fig:ba-accuracy-matrix} for a detailed description of each panel.}
    \label{fig:dmx-accuracy-matrix}
\end{figure}

Figure~\ref{fig:dmx-accuracy-matrix} shows the model classification accuracy for the two duplication divergence models as a heat map for the different methods for feature selection. Each method undergoes a phase transition as the cost regularization parameter $\lambda$ grows. The nearest neighbor classifier can perfectly identify the correct model below the threshold, and the accuracy decreases above the transition. In particular, features that quantify clustering, such as the transitivity or average clustering coefficient, are highly informative but they are not ranked highly for large cost regularization values, leading to performance degradation. Notably, mRMR prioritized noise features after including a single highly informative feature because they have low redundancy. Corresponding figures for the other methods can be found in \supref{Figure~\ref{fig:dmx-accuracy-matrix-appendix} in Appendix~\ref{sup:sec:heatmaps}}{Figure~2 of the online supplement}.

The results using cost-based methods for feature selection are encouraging. For all methods, a penalization parameter value exists that significantly decreases the total feature evaluation cost without affecting the classification accuracy.

\subsection{Feature ranking using smaller pilot networks}
\label{sec:small_networks}

Cost-based feature selection methods can determine a set of inexpensive but informative features. However, applying these methods remains challenging when the cost of obtaining the first reference table for training is extremely high. To address this challenge, we reduced the cost of obtaining the initial reference table by reducing the number of nodes in the simulated networks. We generated pilot simulations comprising networks with $n_s$ nodes instead of $n$ nodes, and we applied the selection methods without cost penalization, i.e., $\lambda=0$. 

We focused on the same two model selection problems as presented in Section \ref{sec:efficiency_cost_based_selection}, and the simulation procedure was unchanged except for the the size of the networks used to generate the first reference table. We varied the size of simulated networks from $n_s=100$ to $n_s=1,000$ in steps of 100 nodes. We evaluated the ability of the selection methods to choose features based on pilot simulations by evaluating the classification accuracy on the same held-out test set with $n=1,000$ nodes presented in Section \ref{sec:efficiency_cost_based_selection}.

\begin{figure}
    \spacingset{1}
    \centering
    \includegraphics{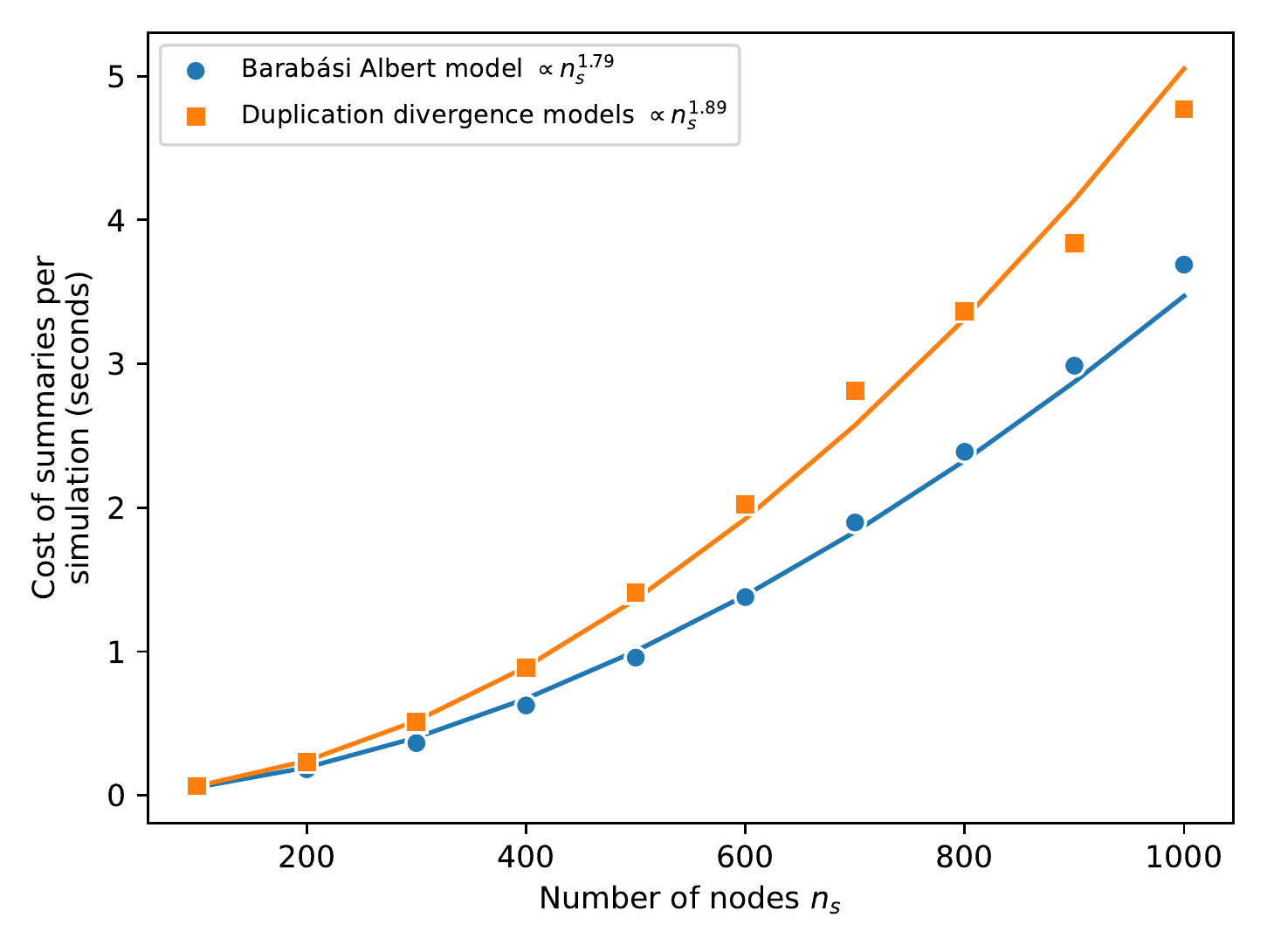}
    \caption{\emph{Computational cost for features grows superlinearly with network size.} Markers indicate the computational cost of evaluating features for one realization of pilot networks of variable size. Lines correspond to monomial fits, demonstrating the superlinear scaling of the computational burden with increasing network size.}
    \label{fig:pilot-simulation-cost}
\end{figure}

As shown in Figure~\ref{fig:pilot-simulation-cost}, the computational cost scaled superlinearly with the number of nodes for both model classes (BA and duplication divergence models). Thus, the overall simulation time can be reduced substantially by considering pilot simulations with a smaller number of nodes. In our experiments, simulating networks with only $n_s=100$ nodes compared with $n_s=1,000$ nodes resulted in a performance gain factor of 57 and 73 for BA and duplication divergence models, respectively. Despite the impressive performance gains for both model classes, the classification accuracy remained unchanged at 100\% for all feature selection methods, provided two or more features were selected. If only a single feature was selected, ReliefF-based approaches performed poorly and RF-based methods exhibited variable performance due to their stochastic nature, as discussed in Section \ref{subsec:BA:cost_based_1000nodes}. These problems are less important in practice, because we typically include more than one feature in analyses to capture complementary information about networks.

\subsection{Application to yeast protein interaction networks}

\begin{figure}
    \spacingset{1}
    \centering
    \includegraphics{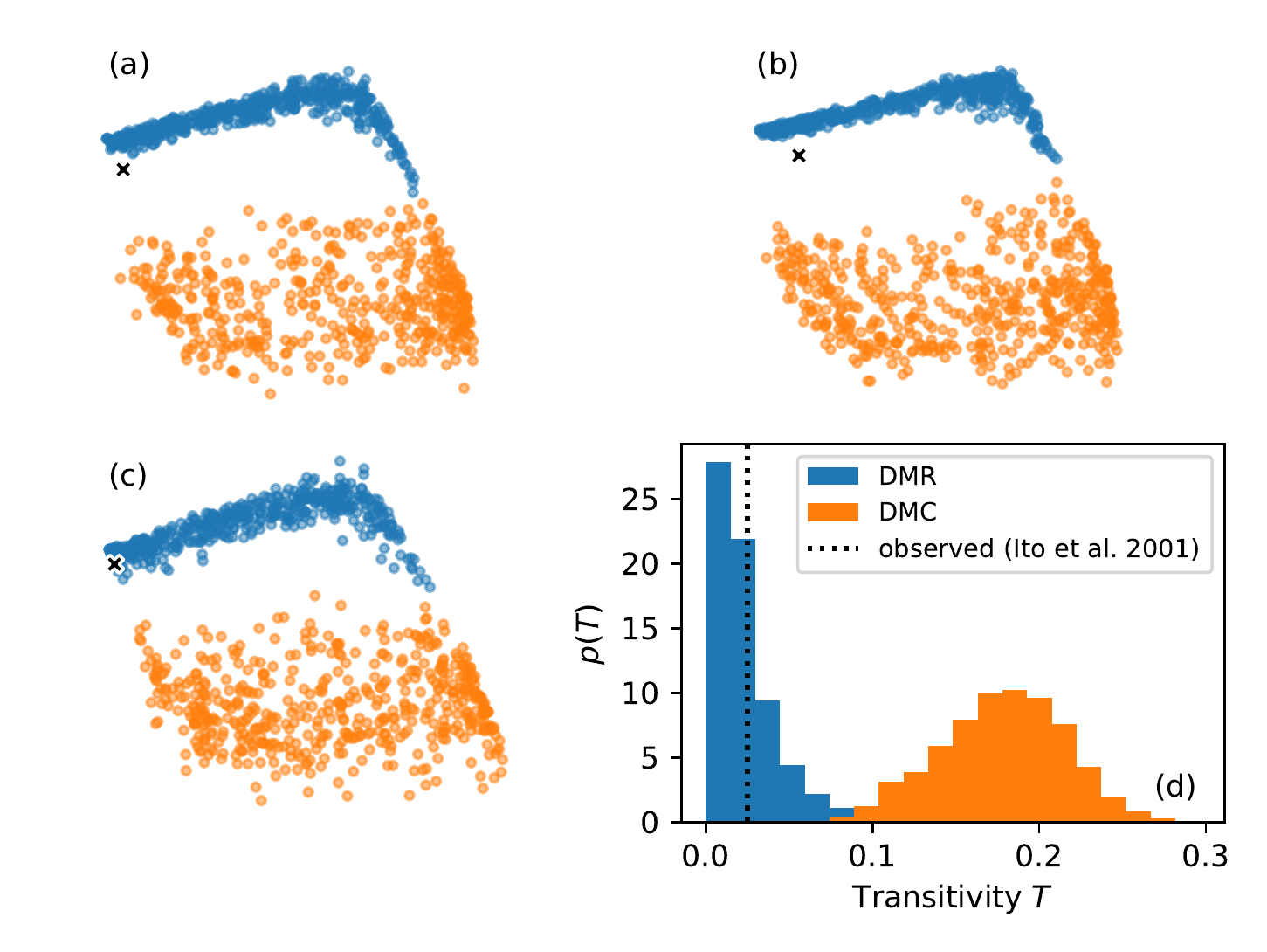}
    \caption{\emph{Three different yeast protein interaction networks are more likely to have been generated by duplication divergence models with random mutation than with complementation.} Panels~(a) through~(c) show embeddings of simulated graphs generated as round markers for the data reported in \citet{ito:2001}, \citet{yu:2008}, and \citet{uetz:2000}, respectively. The cross symbol represents the corresponding observed graph. Embeddings were generated using multidimensional scaling to approximate the Euclidean distance based on the top five ranked features using JMI with a $\lambda=1.0$ cost regularization. Panel~(d) shows the distribution of transitivity values, the highest-ranked feature, for each of the two models. The dashed vertical line corresponds to the transitivity of the protein interaction graph from \citet{ito:2001}.}
    \label{fig:yeast-mds}
\end{figure}

To demonstrate the utility of our cost-based feature selection methods for real-world data, we applied them to three different yeast protein interaction networks. The datasets comprise pairwise physical interactions between pairs of proteins, and we sought to establish which of the two duplication divergence models presented in Section~\ref{subsec:BA:cost_based_1000nodes} provides a better fit for the data. The three interaction networks have $n=437$ \citep{uetz:2000}, $n=813$ \citep{ito:2001}, and $n=1,278$ \citep{yu:2008} nodes. We simulated a reference table of $N=5,000$ networks with the corresponding number of nodes for each network. As demonstrated in Sections~\ref{sec:pilot-simulations} and \ref{sec:cost-based-methods}, respectively, the number of nodes used for feature ranking did not materially affect the classification accuracy, and the classification accuracy was high irrespective of the feature selection method used. In this example, we ranked features based on the reference table of $n_s=1,000$ nodes described in Section~\ref{subsec:DMC_DMR:cost_based_1000nodes} using the JMI method described in Section~\ref{subsubsec:joint-mutual-information} with a cost regularization of $\lambda=1.0$. We used the five highest-ranked features (transitivity, size of the three-core, average clustering coefficient, average local efficiency, and size of the four-core) to train a nearest neighbor classifier after standardizing features, reducing the computational burden by a factor of 20. 

As shown in Figure~\ref{fig:yeast-mds}, all three models strongly preferred the duplication divergence model with random mutations over complementation. The two-dimensional embeddings obtained using multidimensional scaling \citep{Borg1996} closely approximate distances in the five-dimensional feature space (distance correlation $>0.6$ for all datasets) such that neighbors in the embedding space are likely neighbors in the underlying feature space. Panel~(d) shows the prior predictive distribution for the highest ranked feature for discriminating between the two models: the transitivity. The distributions were well separated (two-sample Kolmogorov-Smirnov test statistic $>0.98$ with $p$-value $<10^{-6}$) confirming that the transitivity is a highly-informative feature. The coordinates in feature space corresponding to each dataset were closest to samples from the DMR model; however, these points lay at the boundary of the manifold of simulations \citep{yu:2008} or not on the manifold \citep{ito:2001,uetz:2000}, suggesting that neither model can fully explain the data.

\section{Discussion}
\label{sec:discussion}

Selecting informative features is crucial for model selection tasks. Accounting for the computational burden of features is of practical significance for accelerating inference and shortening the cycle of model iteration. This is especially important for studying mechanistic network models with features that are computationally intensive to evaluate. We showed that cost-based feature selection methods can be used to select features that greatly reduce the computational burden without affecting the classification accuracy. 

While we focused on filter selection methods that provide a ranking of the features, more complex subset exploration strategies could be employed that are not limited to the filter category. The proposal of \citet{zhang:etal:2019}, i.e., a cost-based wrapper algorithm with subset exploration based on an artificial bee colony, would be an interesting method to explore. Moreover, the cost-based literature is sparse and integrating recent feature selection algorithms, such as additional variants of ReliefF, into this cost-based framework would be highly beneficial \citep{urbanowicz:etal:2018:review}.

We also investigated the relevance of features selected using pilot simulations that have fewer nodes than the observed network. For the two network models studied, even reducing the number of nodes by an order of magnitude did not affect the performance of the downstream model selection algorithm. As the computational cost to evaluate features scales superlinearly with network size, this approach can substantially decrease the required resources---at least a factor of 50 in our experiments. Future efforts should establish the limits of this approach, as we know that the method will eventually fail: distinguishing between the two models based on the same seed network is impossible.

One possibility in the context of model choice is to select a value for $n_s$ that gives rise to clusters of features of simulated data such that the clusters are clearly separated based on the model that was used to generate them, as illustrated in panel~(d) of Figure~\ref{fig:pilot-simulation-cost}. While a low $n_s$ value is more likely to lead to networks with similar features, a higher value would be expected to highlight model-specific network features, which should facilitate model identification. Distance between clusters or (dis)similarity of clusters could also be used, and these clusters could be based on the full feature space or on a reduced space obtained, for example, using discriminant analyses.

In this paper, we evaluated the accuracy of nearest neighbor classifiers, and we showed that, for cost-based filter methods for feature selection, the penalization parameter $\lambda$ can be tuned such that the computational cost of generating features is reduced without compromising their predictive performance. Because classic filter methods are classifier independent, features can be selected as a preprocessing step. The selected features can be used for any downstream model selection task---even computationally challenging approaches such as ABC to recover the full posterior.

\section{Supplementary materials}
\label{sec:supplement}

\begin{description}
    \item[python-package for cost-based feature selection]: python-package \texttt{cost\_based\_se\-lec\-tion} contains code to select features based on both their informativeness and (computational) cost. Instructions for reproducing the results presented in this article are available in the README.
\end{description}

\section*{Declarations}

\begin{description}
\item[Funding:] This project was supported by the U.S. National Institutes of Health award R01AI138901.

\item[Conflict of interest:] The authors declare no conflict of interest.

\item[Authors' contributions:] Conceptualization: LR and JPO; Methodology: LR, TH, and JPO; Formal analysis and investigation: LR, TH, and JPO; Writing - original draft preparation: LR; Writing - review and editing: LR, TH, and JPO; Funding acquisition: JPO.

\item[Acknowledgment:] We would like to thank Antonietta Mira for useful discussions about feature selection, Hali Hambridge and Jonathan Larson for their comments and suggestions on the manuscript, and Amanda King for proofreading an early version of this paper.

\end{description}

\bibliographystyle{abbrvnat}
\bibliography{references}

\appendix
\section{Supplementary accuracy heatmaps}
\label{sup:sec:heatmaps}

\begin{figure}
    \spacingset{1}
    \centering
    \includegraphics{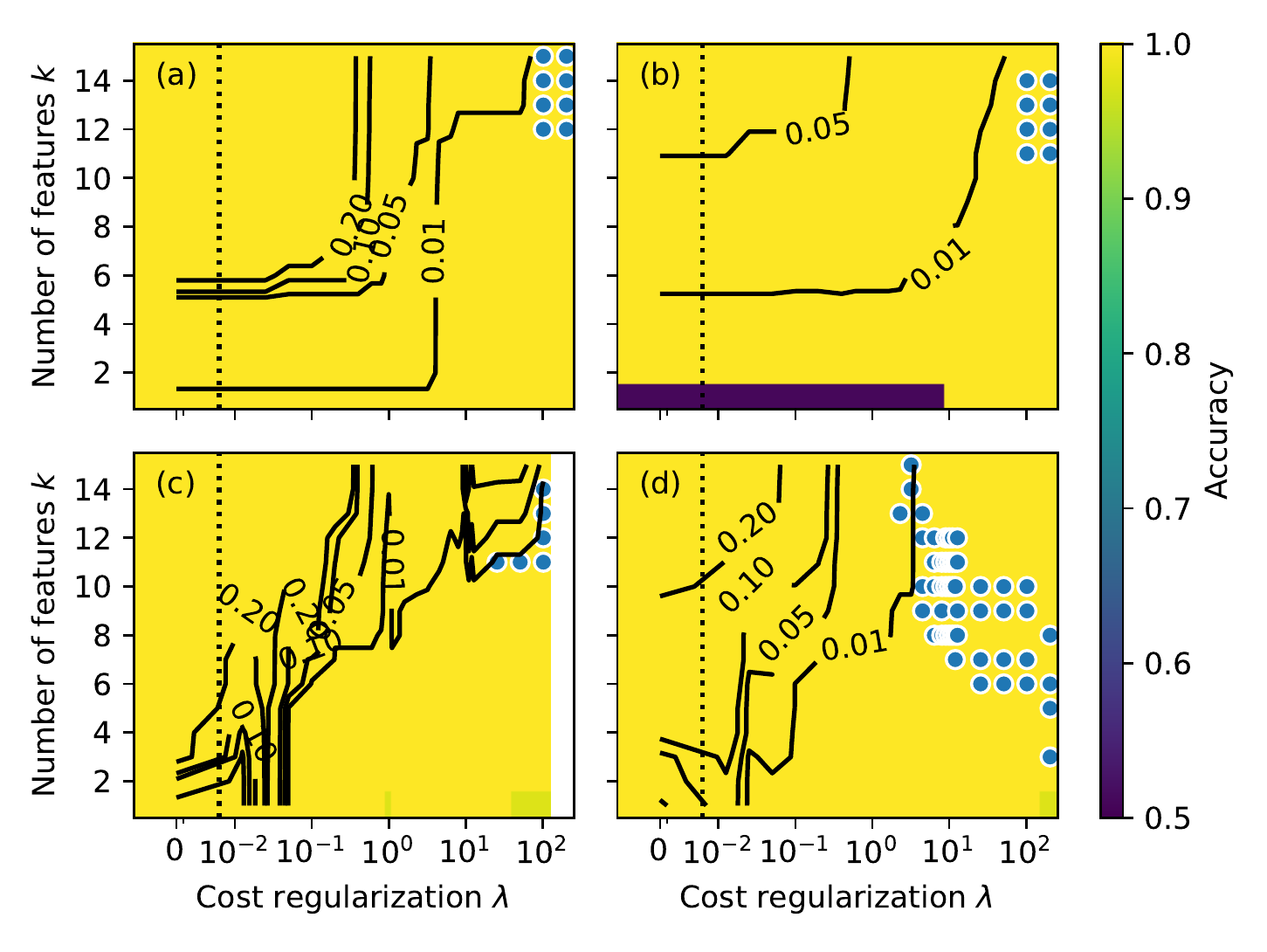}
    \caption{Same as \supref{Figure~\ref{fig:ba-accuracy-matrix}}{Figure~1} in the main text for four different methods in panels~(a) through~(d): JMIM, random-forest-based reliefF, weighted random forest based on mean decreased accuracy, penalized random forest and importance based on impurity. Weighted random forest based on impurity is not shown as it has similar characteristics as weighted random forest based on mean decreased accuracy.}
    \label{fig:ba-accuracy-matrix-appendix}
\end{figure}

\begin{figure}
    \spacingset{1}
    \centering
    \includegraphics{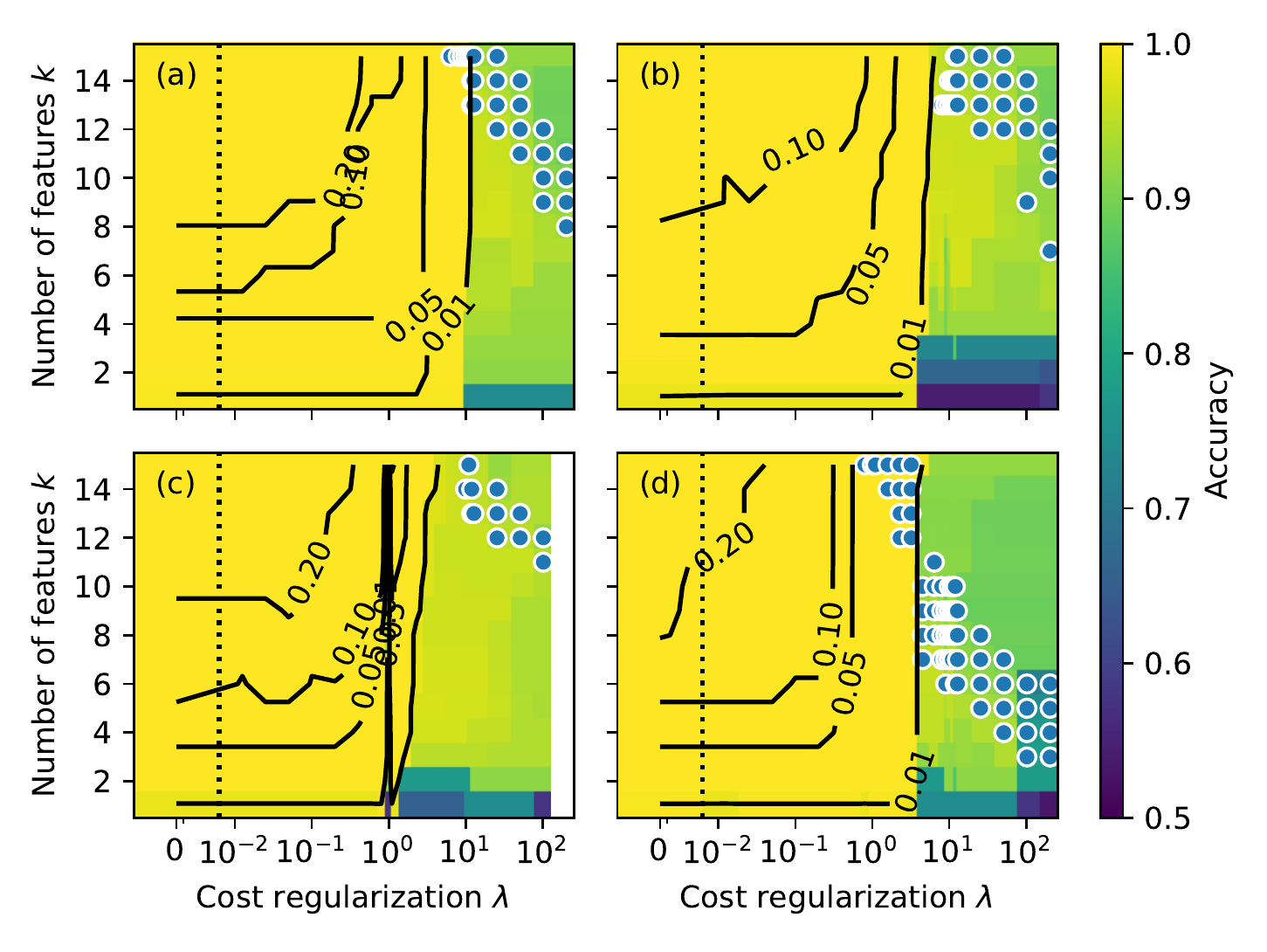}
    \caption{Same as \supref{Figure~\ref{fig:dmx-accuracy-matrix}}{Figure 2} in the main text for four different methods in panels~(a) through~(d): JMIM, random-forest-based reliefF, weighted random forest based on mean decreased accuracy, penalized random forest and importance based on impurity. Weighted random forest based on impurity is not shown as it has similar characteristics as weighted random forest based on mean decreased accuracy.}
    \label{fig:dmx-accuracy-matrix-appendix}
\end{figure}

Figures \ref{fig:ba-accuracy-matrix-appendix} and \ref{fig:dmx-accuracy-matrix-appendix} show further accuracy heatmaps for the Barab\'asi-Albert models and duplication divergence models, respectively.

\section{Network summary statistics}
\label{sup:sec:feature_statistics}

Table \ref{sup:tab:feature-statistics-used} provides the list of the summary statistics used in our analyses including computational costs.

\spacingset{1}
\begin{longtable}{lrr}
\caption{The 51 network summary statistics considered. All can be computed easily with the Python packages \texttt{NetworkX} \citep{hagberg:etal:2008} or \texttt{NetworKit} \citep{networkit}. LCC stands for ``largest connected component'' and DD stands for ``degree distribution.'' Reported runtimes are averaged over 25,000 simulated graphs comprising 1,000 nodes each. Processing summaries for the reference table of $N=5,000$ simulations takes 5.2~h for the Barabasi-Albert models and 6.6~h for the protein interaction network models on a single core. All experiments were run on a 2020 M1 MacBook Pro with 16~GB of memory.}
\label{sup:tab:feature-statistics-used} \\ 

\toprule
 & \multicolumn{2}{c}{\textbf{Mean runtime (milliseconds)}} \\
\cmidrule{2-3}
\textbf{Feature} & \multicolumn{1}{l}{\textbf{Barabasi-Albert}} & \multicolumn{1}{l}{\textbf{Protein Interaction}} \\ \midrule \endfirsthead

\multicolumn{3}{c}%
{{\bfseries \tablename\ \thetable{} -- continued from previous page}} \\
\toprule
 & \multicolumn{2}{c}{\textbf{Mean runtime (milliseconds)}} \\
 \cmidrule{2-3}
\textbf{Feature} & \multicolumn{1}{l}{\textbf{Barabasi-Albert}} & \multicolumn{1}{l}{\textbf{Protein Interaction}}\\\midrule
\endhead

\multicolumn{3}{c}{\dotfill continued on next page \dotfill} \\ \bottomrule
\endfoot

\bottomrule
\endlastfoot

\multicolumn{3}{c}{\textbf{General graph structure}}\\
Number of connected components & 0.7 & 0.7\\
Number of nodes in LCC & 13.0 & 18.4\\
Number of edges in LCC & 13.2 & 18.5\\
\multicolumn{3}{c}{\textbf{Distance measures}}\\
Diameter of the LCC & 712.8 & 448.6\\
Average geodesic distance in LCC & 785.7 & 484.4\\
Average global efficiency & 111.3 & 100.7\\
Average local efficiency in LCC & 42.4 & 92.5\\
\multicolumn{3}{c}{\textbf{Centrality}}\\
Average degree connectivity & 3.9 & 4.3\\
Average degree connectivity in LCC & 16.8 & 22.0\\
Estrada index & 257.6 & 193.7\\
Entropy of the DD & 0.2 & 0.2\\
Maximal degree & 0.2 & 0.2\\
Average degree & 0.2 & 0.2\\
Median degree & 0.2 & 0.2\\
Standard deviation of the DD & 0.2 & 0.2\\
25\% quantile of the DD & 0.3 & 0.3\\
75\% quantile of the DD & 0.3 & 0.2\\
Average betweenness centrality & 324.6 & 431.1\\
Maximal betweenness centrality & 324.5 & 431.1\\
Average eigenvector centrality & $<0.1$ & $<0.1$\\
Maximal eigenvector centrality & $<0.1$ & $<0.1$\\
Central point dominance & 324.6 & 431.2\\
\multicolumn{3}{c}{\textbf{Groups of nodes}}\\
Transitivity & 11.2 & 44.7\\
Number of triangles & 10.5 & 44.6\\
Average clustering coefficient & 10.9 & 44.5\\
Average square clustering & 12.0 & 254.8\\
Median square clustering & 12.0 & 254.8\\
Std. dev. square clustering & 12.0 & 254.8\\
Size of 2-core & 16.0 & 31.9\\
Size of 3-core & 12.5 & 27.9\\
Size of 4-core & 8.6 & 25.2\\
Size of 5-core & 3.9 & 23.1\\
Size of 6-core & 3.9 & 21.6\\
Number of 2-shells & 7.0 & 7.6\\
Number of 3-shells & 7.7 & 7.6\\
Number of 4-shells & 8.5 & 7.4\\
Number of 5-shells & 3.9 & 7.3\\
Number of 6-shells & 3.9 & 7.2\\
Number of 4-cliques & 23.3 & 128.6\\
Number of 5-cliques & 23.2 & 128.6\\
Number of 3-shortest paths & 109.8 & 99.2\\
Number of 4-shortest paths & 109.6 & 99.0\\
Number of 5-shortest paths & 109.6 & 99.0\\
Number of 6-shortest paths & 109.6 & 99.0\\
Maximal clique size & 23.2 & 128.6\\
Size of minimum node dominating set & 132.1 & 243.0\\
Size of minimum edge dominating set & 0.7 & 1.0\\
\multicolumn{3}{c}{\textbf{Synthetic noise}}\\
Standard normal & $<0.1$ & $<0.1$\\
Continuous uniform on $[0,50)$ & $<0.1$ & $<0.1$\\
Bernoulli with success probability 0.5 & $<0.1$ & $<0.1$\\
Discrete uniform on $[0..50)$ & $<0.1$ & $<0.1$\\
\midrule
\textbf{Total} & 3719.0 & 4769.8\\

\end{longtable}
\spacingset{1.5}

\section{ReliefF scores for three-core and mean degree features}
\label{app:three-core}

As discussed in \supref{Section~\ref{par:relief-approaches}}{Section~2.1.3} of the main text, RelieF-based algorithms assign a higher score to the size of the three-core than the mean degree, despite the latter being more informative than the former. Table~\ref{tab:three-core-mean-degree} shows the within- and between-classes in different dimensions, illustrating why the method fails to identify the more informative feature.

\begin{table}
    \spacingset{1}
    \centering
    \begin{tabular}{rrrrrcrrrr}
        \toprule
        \multicolumn{1}{l}{\textbf{Feature}} & \multicolumn{4}{c}{\textbf{three-core size}} && \multicolumn{4}{c}{\textbf{mean degree}} \\
        \cmidrule{2-5} \cmidrule{7-10}
        $\mathbf{m=\ldots}$ & \textbf1 & \textbf2 & \textbf3 & \textbf4 && \textbf1 & \textbf2 & \textbf3 & \textbf4 \\
        \midrule
        \textbf1 & - & 0 & 1 & 1 && -   & 1/3 & 2/3 & 1\\
        \textbf2 & 0 & - & 1 & 1 && 1/3 & -   & 1/3 & 2/3\\
        \textbf3 & 1 & 1 & - & 0 && 2/3 & 1/3 & -   & 1/3\\
        \textbf4 & 1 & 1 & 0 & - && 1   & 2/3 & 1/3 & -\\
        \midrule
        \multirow{ 2}{*}{\textbf{mean dist.}} & 2/3 & 2/3 & 2/3 & 2/3 && 2/3 & 4/9 & 4/9 & 2/3 \\
        \cmidrule{2-5} \cmidrule{7-10}
        &\multicolumn{4}{c}{$2/3=0.666\ldots$}&&\multicolumn{4}{c}{$5/9=0.555\ldots$}\\
        \bottomrule
    \end{tabular}
    \caption{\emph{The mean out-of-class distance for the three-core size is larger than for the mean degree even though the latter is more informative than the former.} The table shows the distance between all combinations of the Barab\'asi-Albert models considered in the main text along two dimensions after range normalization: the three-core size and the mean degree. The second to last row lists the mean out-of-class distance for a given model, and the last row shows the out-of-class distances averaged over all models (assuming a uniform prior over model labels).}
    \label{tab:three-core-mean-degree}
\end{table}

\end{document}